\documentclass[twocolumn,floatfix,superscriptaddress,longbibliography,nofootinbib]{revtex4-1}
\RequirePackage[sort&compress]{natbib}

\usepackage{bbold}
\usepackage{bbm}
\usepackage[pdftex]{graphicx}
\usepackage{latexsym,amsmath,verbatim}
\usepackage{color}
\usepackage{rotating}
\usepackage{multirow}
\usepackage[english]{babel}
\usepackage{color}

\newcommand{\quot}[1]{``#1''}

\newcommand{\Fig}[1]{Fig.\,\ref{#1}}

\newcommand{\heading}[1]{{\vspace{0.25truecm}\noindent\textbf{#1.}}}

\begin{document}

\title{Unraveling the Origin of Social Bursts in Collective Attention}

\author{Manlio De Domenico}
\email[Corresponding author:~]{mdedomenico@fbk.eu}%
\affiliation{Fondazione Bruno Kessler, Via Sommarive 18, 38123 Povo (TN), Italy}
\affiliation{Max Planck Institute for the Physics of Complex Systems, 01187 Dresden, Germany}

\author{Eduardo G. Altmann}
\affiliation{School of Mathematics and Statistics, The University of Sydney, 2006 NSW, Australia}
\affiliation{Max Planck Institute for the Physics of Complex Systems, 01187 Dresden, Germany}

\date{\today}


\begin{abstract} 
In the era of social media, every day billions of individuals produce content in socio-technical systems resulting in a deluge of information. However, human attention is a limited resource and it is increasingly challenging to consume the most suitable content for one's interests. In fact, the complex interplay between individual and social activities in social systems overwhelmed by information results in bursty activity of collective attention which are still poorly understood. Here, we tackle this challenge by analyzing the online activity of millions of users in a popular microblogging platform during exceptional events, from NBA Finals to the elections of Pope Francis and the discovery of gravitational waves. We observe extreme fluctuations in collective attention that we are able to characterize and explain by considering the co-occurrence of two fundamental factors: the heterogeneity of social interactions and the preferential attention towards influential users. Our findings demonstrate how combining simple mechanisms provides a route towards complex social phenomena.
\end{abstract}

\maketitle






The ability to filter the most relevant data out of a deluge of information characterizes human intelligence. When this ability is coupled with individual's behavioral responses, like deciding to take an action based on the processed information, intriguing phenomena~\cite{bagrow2011collective} such as collective attention might emerge. Like popularity, attention depends on a variety of both endogenous and exogenous factors that have effects on several aspects of human behavior, from timing patterns of activity~\cite{barabasi2005origin} to peculiar responses to shocks~\cite{roehner2004response}. The advent of social media and the possibility to record the simultaneous activity of millions of individuals allows the study of this type of phenomena on unprecedented large scales. In fact, such responses are often characterized by information cascades~\cite{lerman2010information,bakshy2011everyone,wu2011says,banos2013role,goel2015structural} and exhibit a rich dynamics with a long memory which is responsible, for instance, for the emergence of power-law distributed physical observables such as waiting times~\cite{crane2008robust,crane2010power} and responses to social-media items~\cite{mitchell2017twitter}. This dynamics has been successfully modeled by a special class of self-exciting point processes known as Hawkes processes~\cite{hawkes1971spectra}, described by a self-reinforced dynamics where the likelihood of future events increases with the occurrence of a specific event.

Like online popularity~\cite{szabo2010predicting,ratkiewicz2010characterizing,borghol2011characterizing,figueiredo2011tube,de2012popularity,bandari2012pulse,pinto2013using}, collective attention is characterized by a quickly growing accumulated focus on a specific topic, e.g. presidential elections discussion on socio-technical systems, until a well identified peak of attention is reached, followed by a phase of decreasing interest with a slow decay~\cite{naaman2011hip,lehmann2012dynamical,omodei2015characterizing}. The dynamical features of both the rise and decline of attention are still debated, although there is some evidence in support of power-law distributed activity~\cite{crane2008robust,crane2010power,ratkiewicz2010characterizing} which is a signature of criticality in complex networked systems~\cite{dorogovtsev2008critical}. On the one hand, some studies succeeded in providing a description of collective attention dynamics while neglecting the effects of the underlying social structure~\cite{wu2007novelty}. In this case, popularity and the rise of attention are understood as the result of an extrinsic factor -- the amount of promotions from the outside world that a content, such as a video or a news, receives -- acting upon two intrinsic factors, namely sensitivity to promotion and inherent virality~\cite{rizoiu2017expecting}. On the other hand, recent studies highlighted the effects of the topological features, i.e. the underlying network of interactions, as well as of competing dynamics and memory time on the spreading phenomena observed in socio-technical systems~\cite{gleeson2016effects}. Along this direction, many studies proposed different models based on the interplay between social structure and complex spreading dynamics to characterize the collective behavior observed in social media~\cite{myers2014bursty}, specially during special events such as the discovery of the ``God particle''~\cite{dedomenico2013anatomy} or in response to real-world exogenous shocks such as disasters~\cite{he2017measuring}. The interplay between system's topology and statistics of exogenous factors -- such as news media -- determine time-dependent network correlations that have been captured by more complex dynamical models of human activity, such as non-stationary~\cite{tannenbaum2017theory} and non-linear~\cite{fujita2018identifying} Hawkes processes and stochastic differential equations with L\'evy noise~\cite{Miotto2017levy}.

Here, we show that by combining two very simple mechanisms characterizing human activity it is possible to reproduce the most salient statistical features of extreme fluctuations~\cite{bouchaud1990} during collective attention in online social systems, without focusing on the evolution of the underlying dynamics.
More specifically, we consider a preferential attachment process, related to individual's neighborhood and social connectivity that characterizes the network topology, and a preferential attention process, a cognitive dynamics related to individual's attention bias towards specific users of the network.


\section*{Results}

\heading{Overview of the data sets} In this work, we analyze the online activity of millions of users posting millions of messages in Twitter, a popular microblogging platform, during some special events. More details about the data are provided in Tab.~\ref{tab:data}. We focus on events of wide public interest spanning different topics, such as the elections of Pope Francis (religion), NBA finals (sport), the discovery of gravitational waves (science), the Cannes Film Festival (culture). The data sets consist of the second-by-second online activity that for the subsequent analysis has been aggregated at the time scale of $T=1$~min. 

\begin{figure}
\centering
\includegraphics[width=0.49\textwidth]{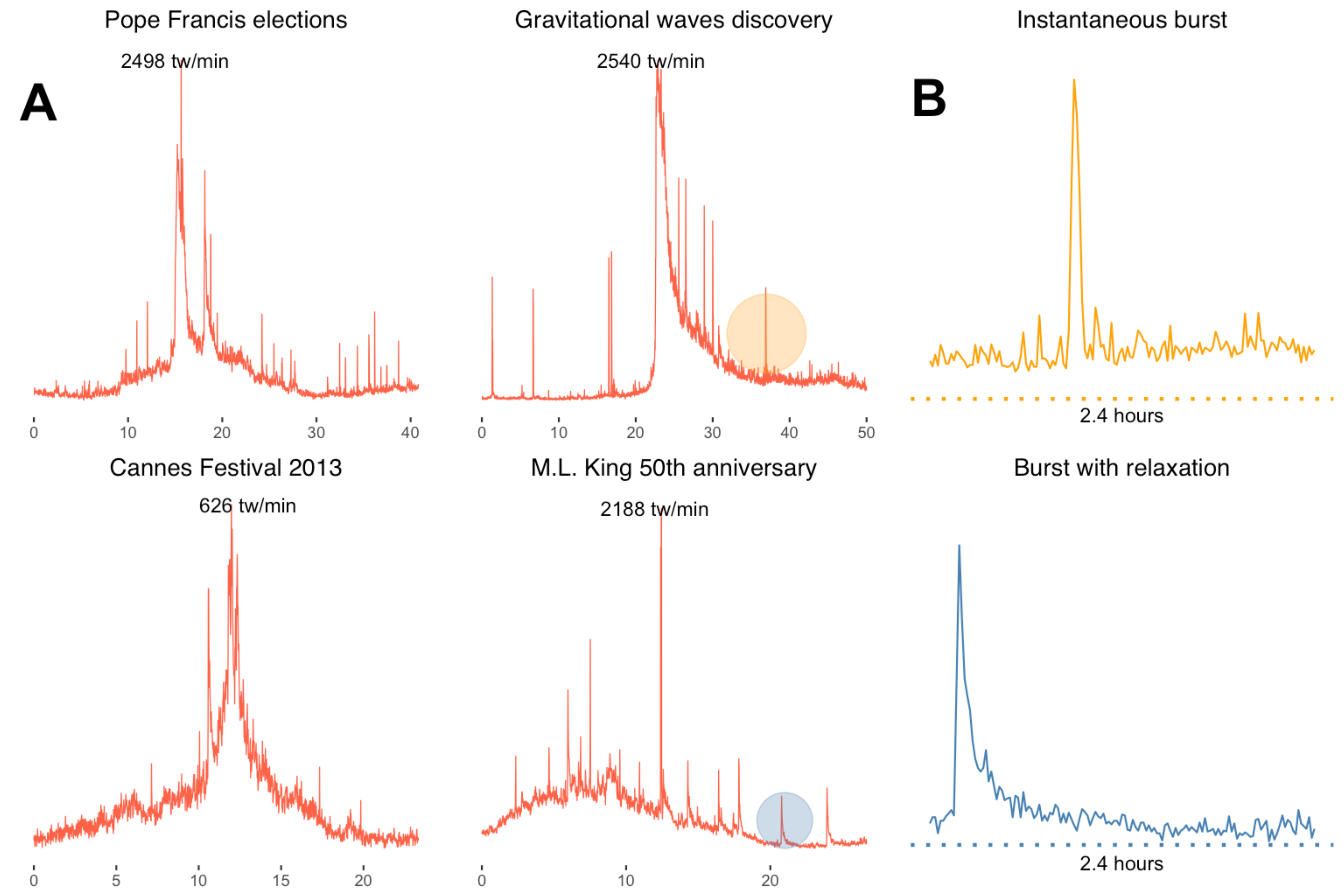}
\caption{\label{fig:Fig1}{\bf Social bursts of collective attention during exceptional events.} (A) Volume of activity in tweets/minute across several hours observed in the microblogging platform Twitter and measured during special events like Pope Francis' election (2013), Cannes Film Festival (2013),  50th anniversary of Martin Luther King's most famous speech (2013) and gravitational waves discovery (2016). (B) Bursts decay either instantaneously (bottom) or with some characteristic relaxation dynamics (top). The collective activity shown here aggregates the number of messages and the social actions they trigger: $N(t)+R_{retweets}(t)+R_{replies}(t)$.}
\end{figure}

\heading{Analysis of bursty activity due to collective attention} Let us focus our attention on the evolution of collective attention dynamics over time, during different special events, shown in \Fig{fig:Fig1}. A common feature, observed in all events regardless of their type (e.g., political, religious, cultural, scientific), is the bursty behavior of the social system: spikes of activity appear to be randomly placed on top of a more smooth temporal variation. \Fig{fig:Fig1}B shows that the spikes are extremely sharp in time, characterized either by an abrupt increase followed by either a decrease of activity within one time unit (1 minute, in the figure) or by a slightly slower decrease of activity resembling the relaxation of a system's response to some stimulus. 

\begin{figure}[!t]
\centering
\includegraphics[width=0.49\textwidth]{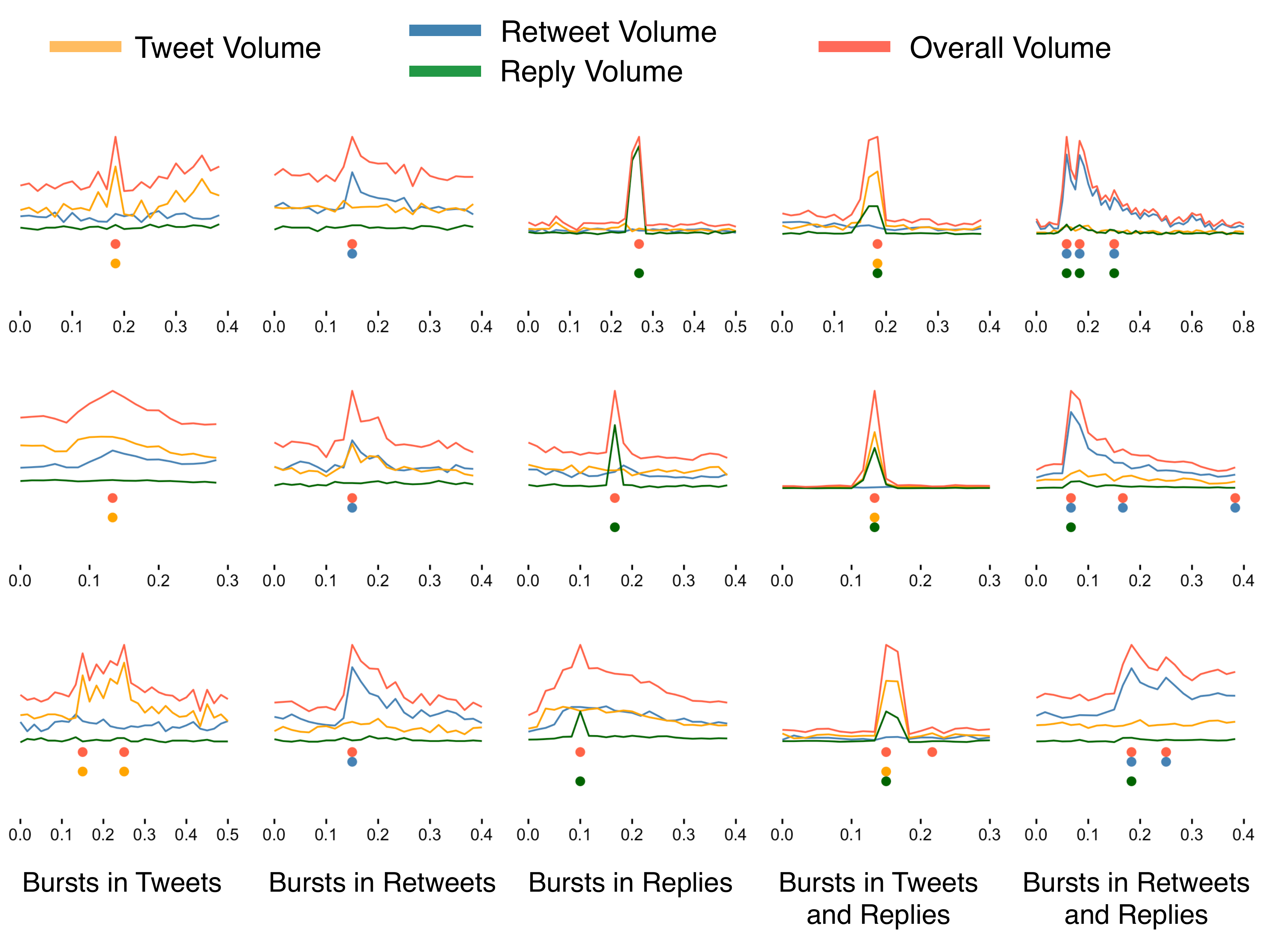}
\caption{\label{fig:Fig2}{\bf Demultiplexing collective attention into specific activities.} Collective attention resulting from the superposition of different individual and social activities (left to right), during different exceptional events (top to bottom). Bursts in the overall activity over time, automatically identified and marked with dots, are mostly driven by social interactions -- Retweet and Reply in this study (see Suppl. Inf. for further analysis) -- which are influenced by the social network's structure.}
\end{figure}

To better understand the nature of such a bursty behavior, we decompose the overall activity into its components due to individual's lone activity (``Tweet'') -- posting messages related to the event which do not involve other users -- and to social interactions, such as endorsing (``Retweet'') or replying to (``Reply'') other individual's posts. \Fig{fig:Fig2} shows that bursts dominated by both individual and social activities exist. Counting the contribution of each activity to many different bursts, compared against random expectations, reveals that the social activities are more frequently responsible for the spikes (see Suppl. Fig.~1--2). The goal of this manuscript is to provide a statistical characterization of this bursty activity and to discuss possible mechanisms that account for them.

\heading{Characterizing bursty activity in collective attention} Our goal is to clarify what type of mechanisms can be responsible for the spiky online activity summarized above. Recent studies attempted to relate the overall collective activity to peculiar characteristics of the underlying social structure or the influence of endogenous and exogenous factors~\cite{fujita2018identifying}. The extremely fast and socially-dominated nature of spikes point towards a mechanism of reinforcement of collective behaviour taking place endogenously in a social network. Our hypothesis is that the variety of fluctuations observed in empirical data are due to the interplay between topological effects, related to the individual's neighborhood and social connectivity, and cognitive effects, related to the individual's bias towards activity from specific users participating into the process. Both effects are known to concentrate the attention in the few most connected users. This motivates us to search for mathematical models that account for the spiky collective attention observed in online platforms such as Twitter and that just depend on individual's relationships and interactions.

We concentrate on the typical case of spikes generated by social activities in response to previous messages. Once a message $i$ is posted, the $k_{i}$ followers of the source user who posted it can act socially (i.e., in Twitter this might correspond to a Reply or a Retweet, corresponding to a direct comment or an endorsement, respectively).  In our simple model, the multiple factors affecting this response are reduced to two: $p_A(t)$ the probability of a follower being active and $q_{i}(t)$ the probability of an active follower to react. The extremely short time scales of the spikes suggests that the reactions to a message are dominated by the immediate followers of the source user, instead of long/deep cascades of interactions in the network. With this simplifying assumptions, the probability that the message $i$ triggers $R_i$ social activities (responses) at time $t$ is given by
\begin{equation}
\label{eq.B}
P(R_i(t)|i) =  B(k_i p_A(t), q_{i}(t)),
\end{equation}
where $B$ is the Binomial distribution with $k_i p_A(t)$ samples and probability $q_{i}(t)$.  The overall social activities $R(t)$ at time $t$ is obtained summing the number of triggered responses $R_i(t)$ over all $N(t)$ messages  contributing to social activities at time $t$ as
\begin{eqnarray}
\label{eq.R} 
R(t)  = \sum_{i=1}^{N(t)} R_i(t)  \approx  p_A(t) \sum_{i=1}^{N(t)} q_{i}(t) k_{i},
\end{eqnarray}
where the approximation is based on the expected number of interactions $k_{i} p_A(t) q_i(t)$, the average of distribution in Eq.~(\ref{eq.B}). In Eq.~(\ref{eq.R}), we consider messages to be randomly placed in the network so that for each message the user associated to $i$  (with $k_i$ and $q_i$) is randomly chosen. In particular, we consider $k_i$ to be a random sample of the degree distribution of the network $\rho(k)$.  Our analysis of empirical data reveals that the duration of bursts due to social actions is, on average, shorter than 5 minutes, with 15 minutes as an upper bound (see Suppl.~Fig.~2). Due to this extremely short time scales of the duration of the bursts, and similar short time scales for the social reactions to posted messages (see Suppl.~Fig.~3), in our model we estimate $N(t)$ (the number of messages contributing to social activities at time $t$) simply as the average number of messages published in a window of time around $t$ (see Ref.~\cite{mitchell2017twitter} for a more detailed account of the slow temporal decay of the number of social interactions to a message). 
It is worth remarking that messages should not be necessarily produced at time $t$, but they can be posted before without triggering social interactions before time $t$.

Equation\,(\ref{eq.R}) defines our simple model for collective attention, and different scenarios are obtained by specifying the network (its degree distribution $\rho(k)$) $p_{A}(t),$ and $q_{i}(t)$. The probability of a users to be active $p_A(t)$ simply re-scales the number $R(t)$ of social activities and will thus not be relevant in our explanation of the spikes. The two critical parameter in the different scenarios are $k_i$ -- the number of users that receive the message $i$ --  and  $q_i(t)$ -- the probability of user $i$ to act socially (retweet/reply). We consider three different scenarios of increasing complexity: 

\begin{itemize}

\item[1.] Homogeneous: $q_i(t)=q(t)$ is independent of $i$ and $\rho(k)$ is sharply peaked around an average degree $\langle k \rangle$ (e.g., $\rho(k)\sim Pois(\langle k \rangle)$). In this case, the role of $q(t)$ is to simply re-scale $p_A(t)$, which are both assumed to have a smooth temporal dependence not related to the spikes. Fluctuations in this scenario are expected to be small because of the well-behaved degree distribution $\rho(k)$, so that this scenario acts as a null model.

\item[2.] Heterogenous: we incorporate to the previous scenario the well-known fact that $\rho(k)$ is a fat-tailed distribution, decaying as $\rho(k)\sim k^{-(1+\mu)}$ for $k\gg 1$.  Typically $1<\mu<2$ and in the specific case of Twitter,  $\mu\simeq 1.2$ was measured~\cite{kwak2010twitter}. Much larger fluctuations are expected in this scenario because of the strong variations in $k_i$ for different $i$, i.e., the messages coming from hubs ($k_i \gg \langle k \rangle$) are expected to receive much more interactions than messages from typical nodes ($k_i \approx \langle k \rangle$).

\item[3.] Preferential attention: we incorporate to the previous scenario the fact that reaction to a message is more likely if it comes from a user that is perceived as important or central. The simplest proxy for such an importance is the degree of the message creator and thus we use $q_{i}(t)\propto k_{i}$. 
\end{itemize}

For each of the scenarios, the sum in Eq.~(\ref{eq.R}) effectively considers samples of distribution with short (case 1) or fat (case 2 and 3) tails. The restriction $2<\mu<3$, valid for all degree distributions, ensures that $\langle k \rangle$ is well defined in scenario 2. In contrast, scenario 3 effectively corresponds to drawing samples from a distribution with exponent $\mu-1$ and therefore with a diverging mean (see Materials and Methods, Model with preferential attention).

\begin{figure*}
\centering
\includegraphics[width=\textwidth]{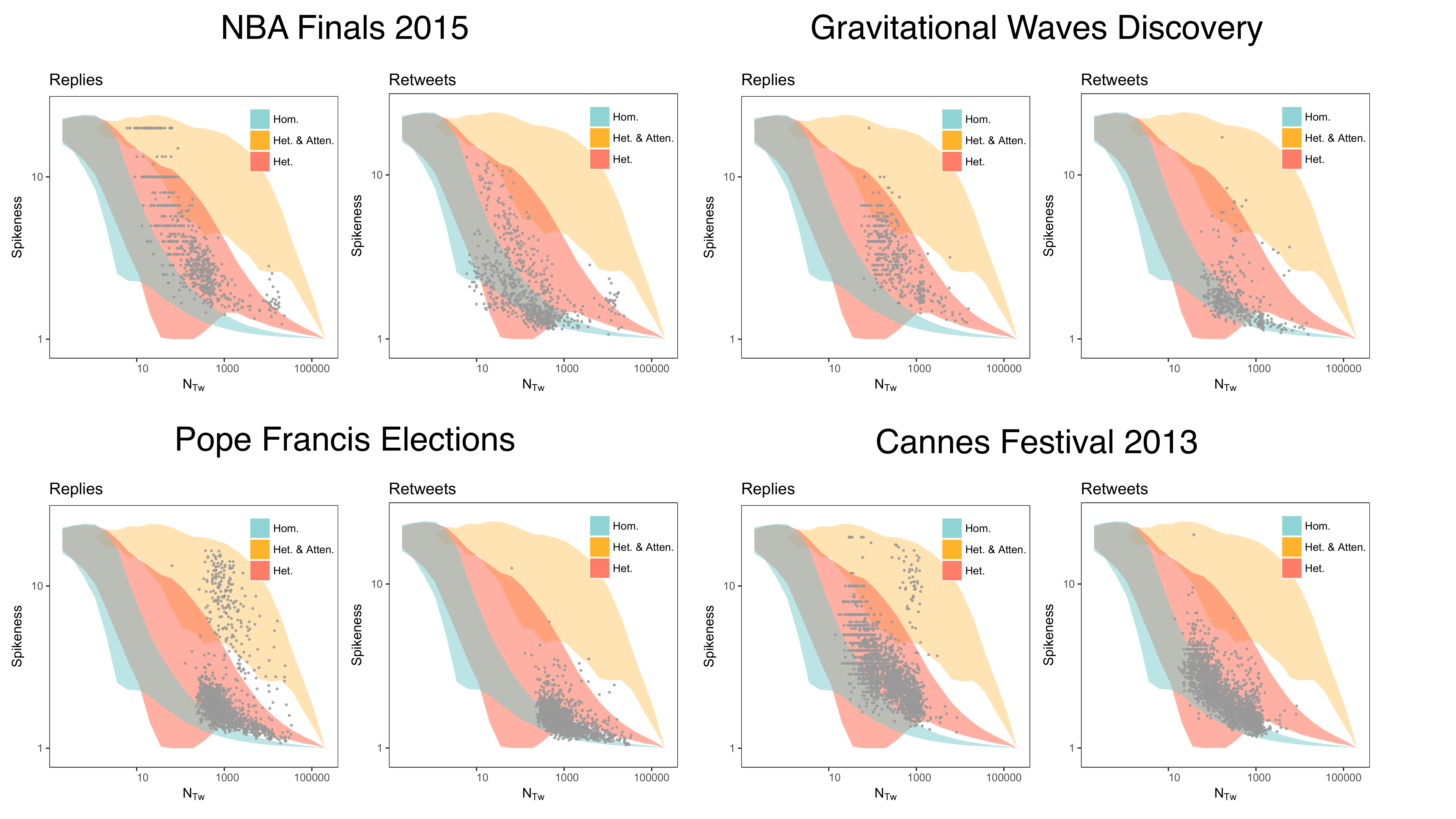}
\caption{\label{fig:Fig3}{\bf Fluctuation analysis of social bursts during collective attention.} Scaling of fluctuations versus number of tweets ($N_{T,w}$) for social activities (replies and retweets) during different exceptional events, for the time window $\ell=20$~minutes (see the text for details). For instance, $N_{T,w}=1000$ indicates an average of 50 posts per minute. Points denote empirical data, whereas shaded areas indicate the 90\% confidence on expectation obtained from simulated collective attention corresponding to i) homogenous social structure with uniformly distributed attention (``Hom.''); ii) social structure obtained from preferential attachment with uniformly distributed attention (``Het.''); iii) social structure obtained from preferential attachment with preferential attention (``Het. \& Atten.'').}
\end{figure*}

\heading{Revealing the mechanisms behind collective attention} 
The mechanisms behind collective attention can be revealed by testing to which extent the scenarios above describe the observations. We are interested in the spikes observed in the data, an extreme case of variability of the activity. Here, the data is represented by a time series of length $L$ encoding the collective activity of the social network over time. We divide this time series into non-overlapping windows of size $\ell$ and, for each window $w=1,2,...,L/\ell$, we quantify the spikiness $S_w$ in the window as the ratio between the maximum and the mean volume $R(t)$ of social responses, for $t$ in the window:
\begin{eqnarray}\label{eq.S}
S_w = \dfrac{\max\limits_{t\in w} R (t)}{\langle R(t) \rangle_{t\in w}}.
\end{eqnarray}
The overall number of posted messages in each window is indicated by  $N_{T,w} = \sum_{t \in w} N(t)$ and we consider $N(t)$ in Eq.~(\ref{eq.R}). First we discuss the expectations for the dependence of $S_w$ on $N_{T,w}$ for the three scenarios of our model. As argued above, the scenarios correspond to random sampling of three fundamentally different types of distribution (short, heavy, and extremely heavy-tails). Accordingly, the $\mathbb{E}[R]$ and $R^{\text{Max}}$ -- the expected value of $R(t)$ and its largest value in $\ell$ independent realizations, respectively -- scale differently with the number of messages $N_{T,w}$, leading to the following estimations of the spikiness $S_w$ (see Materials and Methods, Sum of fat-tailed variables and Maxima):

\begin{itemize}

\item[1.] Homogeneous: $S_w \sim 1 / \sqrt{N_{T,w}}$, i.e., the usual central-limit-theorem decay (i.e., spikiness is not expected for large values of $N_{T,w}$)

\item[2.] Heterogeneous: $S_w \sim 1 /  N_{T,w}^{1-1/\mu}$, i.e., a slower decay of $S_w$ (i.e., spikiness persist for larger values of $N_{T,w}$)

\item[3.] Preferential attention: $S_w$ does not depend on $N_{T,w}$ or, at most, decays slower than algebraic.

\end{itemize}
The scaling (``$\sim$'') relationships above hold for $N_{T,w}(t)\gg 1$, the usual setting of the generalized central limit theorem (see Materials and Methods). When $N_{T,w}(t)\approx 1$, $R(t)$ will follow the distribution of $q_{i}(t)k_{i}$. As anticipated, the activation probability $p_{A}(t)$ just rescales the triggered social activities $R(t)$ in Eq.~(\ref{eq.R}) and therefore it cancels out in the ratio defining $S_w$ in Eq.~(\ref{eq.S}). 

\begin{small}
\begin{table*}[!ht]
\caption{\label{tab:data}Information about data sets used in this study.}
\centering
\begin{tabular}{p{3.5cm}p{1.5cm}p{1.5cm}p{1.5cm}p{1.5cm}p{1cm}p{4.5cm}}
\hline
\textbf{Event} & \textbf{Messages} & \textbf{Users} & \textbf{From} & \textbf{To} & \textbf{Days} &  \textbf{Keywords} \\\hline
Papal Conclave (Pope Francis) & 5,538,257 & 2,530,554 & 2013-02-25 & 2013-03-19 & 21.43 & 
\quot{pope}, \quot{benedict}, \quot{pontifex}, \quot{resign}, \quot{conclave}, \quot{vatican}\\
NBA Finals & 2,993,898 & 1,115,981 & 2015-06-05 & 2015-06-21 & 15.55 & \quot{\#nbafinals}\\
Cannes Film Festival & 1,521,977 & 521,151 & 2013-05-06 & 2013-06-03 & 27.93 & \quot{cannes film festival}, \quot{cannes}, \quot{\#cannes2013}, \quot{\#festivalcannes}, \quot{\#palmdor}, \quot{canneslive}\\
Gravitational Waves Discovery & 859,585 & 451,739 & 2016-02-10 & 2016-02-16 & 5.71 & \quot{\@ligo}, \quot{\#gravitationalwaves}, \quot{\#ligo}, \quot{gravitational waves}, \quot{\#gravitational waves}, \quot{gravitational \#waves}, \quot{onde gravitazionali}, \quot{\#OndesGravitationnelles}, \quot{Ondas gravitacionales}, \quot{Ondes Gravitationnelles}, \quot{\#ondas \#gravitacionales}, \quot{\#ondas gravitacionales}\\
50th Anniv. of M.L. King's ``I have a dream'' speech & 496,094 & 391,467 & 2013-08-25 & 2013-09-02 & 7.77 &
\quot{Martin Luther King}, \quot{\#ihaveadream} \\\hline
\end{tabular}
\end{table*}
\end{small}

In the analysis of the empirical data, typical choices for $\ell$ range from 20~minutes to a few hours: it can not be too small or too large, to allow for a significant number of samples to be analyzed. Each time window consists of $\ell$ measurements, because we have built the data sets at 1~minute resolution.

In order to allow for a meaningful comparison between the data and the results obtained from the model, we generate several independent Monte Carlo realizations of the overall collective activity -- including posting messages and social responses -- in a window of size $\ell$ and, for each realization, we calculate the corresponding spikiness. This is done for increasing values of $N_{T,w}$ and using $N(t) = N_{T,w}/\ell$. This choice of $N(t)$ is justified by the short time scales of the reactions to tweets -- as argued after Eq.~(\ref{eq.R})-- and can be viewed as a lower bound on the number of messages actively generating reactions at time $t$. The value of the spikiness expected from the models and its corresponding variability, respectively $\langle S_{w}\rangle$ and $\sigma_{S_{w}}$, are calculated over the ensemble of Monte Carlo realizations. For each scenario, we build the 90\% confidence interval around the expected spikiness and we evaluate whether the pairs $(N_{T,w}, S_{w})$ measured from the data lie within this region. The results of our analysis are shown in \Fig{fig:Fig3}.
The fluctuation analysis reveals some remarkable features of collective attention. The three models introduced in this work account, all together, for observations. The statistics of social activities can widely vary within the same event, as in the case of Pope Francis election, where replies fluctuations are well explained either by the preferential attachment or the preferential attention models.
In general, the spikes of Retweets are compatible with the heterogeneous model, while the spikes of Replies are larger than expected by the heterogeneous model and can only be accounted in the preferential attention scenario.

\section*{Discussion}

During events of special relevance, collective activity is usually more frenetic -- i.e. the probability of posting is sufficiently high to guarantee a larger number of messages posted to the social system, typically well above 50 messages per minute -- and the overall interest in the subject is driven by external factors. On top of this (smooth) overall tendency, extremely large fluctuations can be observed in form of spikes of activities. This spikes have very short duration (often less than a 1 minute) and reflect a burst of activity and a dramatic concentration of the total social attention. Our main empirical finding is to identify and characterize these spikes. In particular, while spikes can have different origins, most spikes originate from social activities -- such as Replies or Retweets in Twitter -- in response to messages coming from well connected nodes.

We proposed a simple stochastic model to understand the extreme fluctuations observed in social bursts of collective attention. It incorporates two fundamental mechanisms: the preferential attachment process, related to individual's neighborhood and social connectivity that characterize the observed network topology, and a preferential attention process, a cognitive dynamics related to individual's attention bias towards specific users of the network. In this work we considered an heterogeneous connectivity distribution scaling as $k^{-2.2}$, according to independent measurements~\cite{kwak2010twitter}, and attention bias linearly proportional to the connectivity $k$. 
Comparing the model predictions with Twitter data, we find that the more extreme bursts of collective behavior -- typically in form of Replies -- can be understood only through the combination of those two processes.

Our results show that two simple mechanisms are able to reproduce the statistical features of the appearance of spikes during exceptional events and our approach provides a procedure to measure the existence and the influence of preferential attention during events triggering collective attention.


\section*{Materials and Methods}

\heading{Overview of the data sets} Messages posted by users in Twitter, a popular microblogging platform, have been collected using the streaming real-time provided by Twitter API platform, filtered by the specific keywords reported in Tab.~\ref{tab:data}. By default, Twitter  limits to 1\% of the overall number of messages per second that can be retrieved from the streaming API. However, when the fraction of tweets concerning specific keywords is smaller than 1\% of the global volume, Twitter does not apply limitations and the complete flow of information is collected. When this is not the case, Twitter provides messages of warning, reporting the cumulative number of missed tweets. For all events considered in this work, the estimated completeness of the sample is above 95\%. Because of Twitter policies, the data sets (original tweet IDs) are available upon request. Time series, the main data analyzed in this work, will be made publicly available at this URL: www.

\heading{Model with preferential attention} We consider that the probability of
a social action is not a constant $q$ but depends on the message $i$ as $q_{i} = q k_i^\alpha$. From Eq.~(\ref{eq.R}) we obtain 
$$ R_{N} = p_A \sum_{i=1}^{N} q_{i} k_i  \propto \sum_{i=1}^{N} k_i^{1+\alpha}. $$
This can be viewed as a sum of $N$ i.i.d. random variables $x_i \equiv k_i^{1+\alpha}$ which then, a part from pre-factors, is in the form of Eq.~(\ref{eq.sum}). The
distribution of $x_i$, $\rho(x)$, is related to $\rho(k)$ by $ \rho(x) = \rho(k) / dx/dk$. 
For the power law case $\rho(k) \sim k^{-(1+\mu)}$ we obtain 
$$\rho(x) \sim k^{-(1+\mu+\alpha)} \sim x^{-(1+\mu+\alpha)/(1+\alpha))}.$$
This means that the model with preferential attention is equivalent to the model with heterogeneous networks with a modified exponent, obtained by the mapping of the exponents $1+\mu \mapsto 1+\mu'=
(1+\mu+\alpha)/(1+\alpha)$. In the text we consider the case $\alpha=1$ which leads to a modified exponent $\mu' = \mu/2$. In particular, $1<\mu<2$ is mapped to $0.5 < \mu' <1$. The degree distribution of networks is $\mu > 1$ and therefore an effective exponent with $\mu<1$ (the third case discussed below, $0 < \mu < 1$) can only be achieved through the incorporation of preferential attention.

\heading{Sum of fat-tailed variables}
Let $x \le 0$ be a random variable with distribution $\rho(x)$ such that $\rho(x) \sim x^{-(\mu+1)}$ for large $x$, with
$\mu>0$ (fat tails). We are interested in the sum of $N$ independent samples of $x$
\begin{equation}\label{eq.sum}
R_{N} = \sum_{i=1}^{N} x_i.
\end{equation}
Following Ref.~\cite{bouchaud1990},  the following cases can be described:

\begin{itemize}
  \item[1.] \heading{$\mu\ge2$} In this case, which includes also distributions $\rho(x)$ with short tails (such as the Poisson distribution in scenario 1.), both moments $\langle x \rangle $ and $\langle x^2 \rangle$ exist and for large $N$ the usual central limit theorem applies such that $\mathbb{E}[R_{N}]= \langle x \rangle N$ and
    variance $\mathbb{V}[R_{N}] = 2 \sigma_x^2 N$. Therefore, fluctuations are small and decay with $N$ as

    $$\frac{\sqrt{\mathbb{V}[R_{N}]}}{\mathbb{E}[R_{N}]} \sim \frac{1}{\sqrt{N}}.$$

\item[2.] \heading{$ 1 < \mu < 2$} In this case, $\langle x \rangle$ exists but $\langle x^2
  \rangle$ does not. The expected value $\mathbb{E}[R_{N}]= \langle x \rangle N$ holds,
  but the fluctuations increase dramatically. In particular, $\mathbb{V}[R_{N}]$
  diverges with $N$ as
\begin{equation}
(R_{N}-\mathbb{E}[R_{N}])^2 \sim N^{2/\mu},
\end{equation}
and therefore
\begin{equation}
\dfrac{\sqrt{(R_{N}-\mathbb{E}[R_{N}])^2}}{\mathbb{E}(R_{N})} \sim
N^{(1-\mu)/\mu}.
\end{equation}
This still decays to zero because $(1-\mu)/\mu < 0$.

\item[3.] \heading{$ 0 < \mu < 1$} 
In this case $\langle x \rangle$ is not defined and
\begin{equation}
\mathbb{E}[R_{N}] \sim  N^{1/\mu}.
\end{equation}
\end{itemize}

\heading{Maxima} Our measure of spikiness $S_w$ defined in Eq.~(\ref{eq.S}) consider the block-$\ell$ maximum of $R$, denoted by $R^{\text{Max}}$ (i.e., the largest value of $R(t)$ in $\ell$ independent realizations). For $\rho(x) \sim x^{-(1+\mu)}$, the tails of the distribution of $R(t)$ behave as the tails of $\rho(x)$ and therefore, from extreme value theory, we expect the scaling
\begin{equation}
R^{\text{max}} \sim N^{1/\mu},
\end{equation}
for $0<\mu<2$ and $R^{\text{max}} \sim \sqrt{N}$ for $\mu>2$ (including the Poisson distribution).


\section*{Acknowledgments}
M.D.D. acknowledges partial financial support from the Max Planck Institute for the Physics of Complex Systems.  EGA was funded by the University of Sydney bridging Grant G199768.

\section*{Competing financial interests}
The authors declare no competing financial interests.

\bibliographystyle{apsrev4-1}
\bibliography{collective}

\appendix


\renewcommand{\figurename}{Supplementary Figure}
\renewcommand\thefigure{\arabic{figure}}    
\setcounter{figure}{0}    

\begin{figure*}
\centering
\includegraphics[width=12cm]{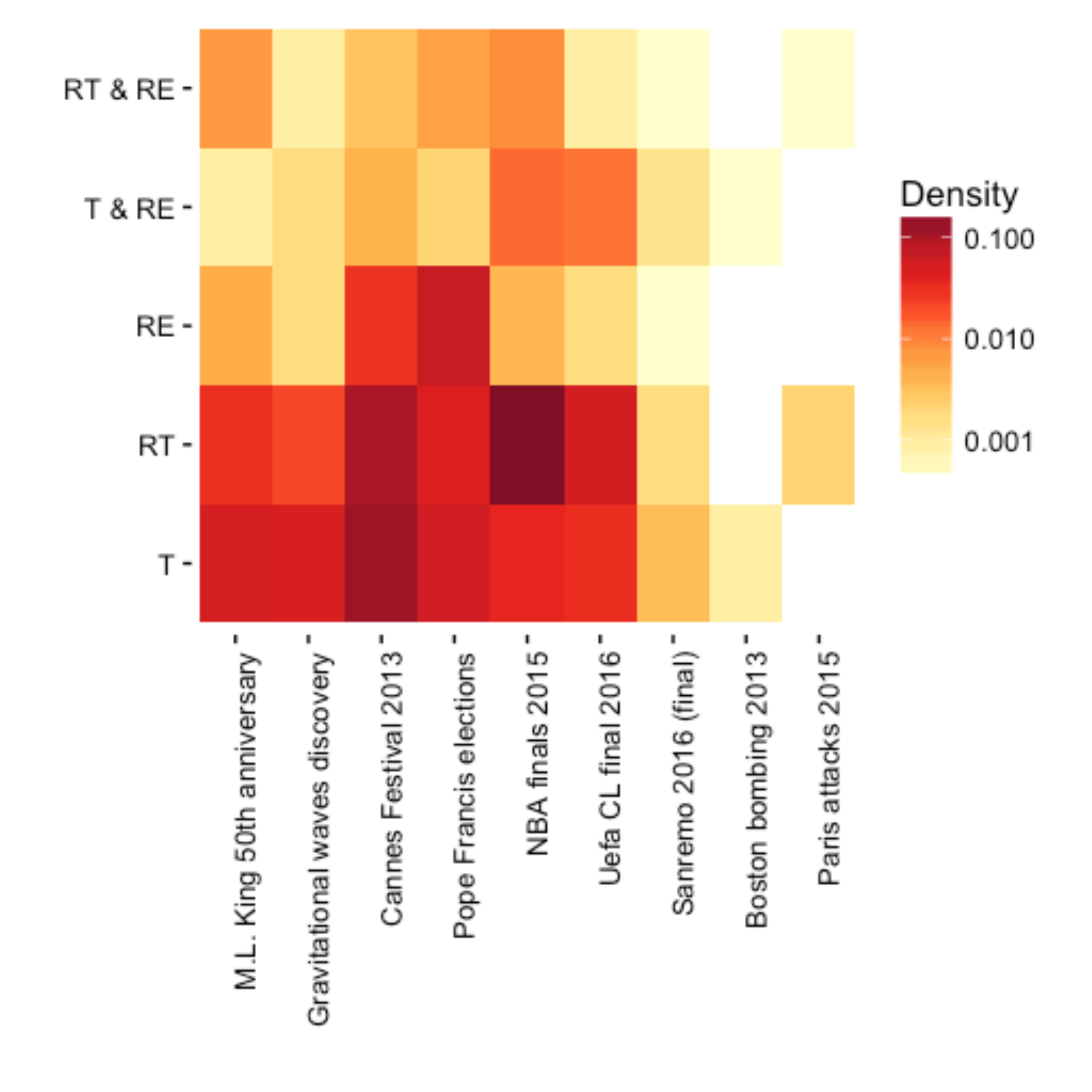}
\caption{{\bf Density of social bursts.} Fraction of bursty activity due to specific actions (T = tweet, RT = Retweet, RE = Reply) and their combinations during 9 exceptional events (see Main Text for an overview of the data sets).}
\label{fig:demultiplex}
\end{figure*}

\begin{figure*}
\centering
\includegraphics[width=12cm]{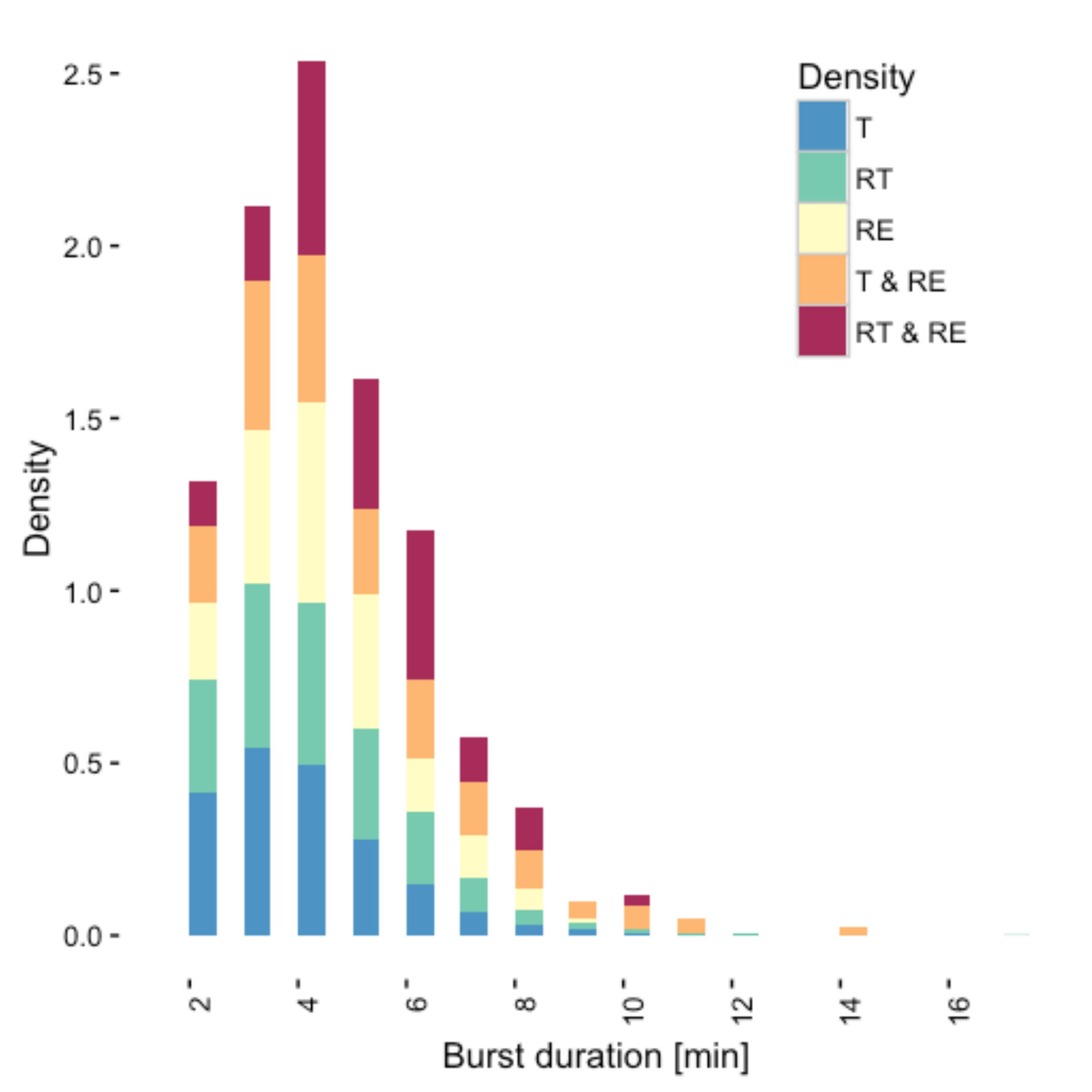}
\caption{{\bf Duration of social bursts.} Distribution of bursts' duration due to specific actions (T = tweet, RT = Retweet, RE = Reply) and their combinations measured from all the exceptional events considered in this study (see Main Text for an overview of the data sets).}
\label{fig:demultiplex}
\end{figure*}

\begin{figure*}
\centering
\includegraphics[width=16cm]{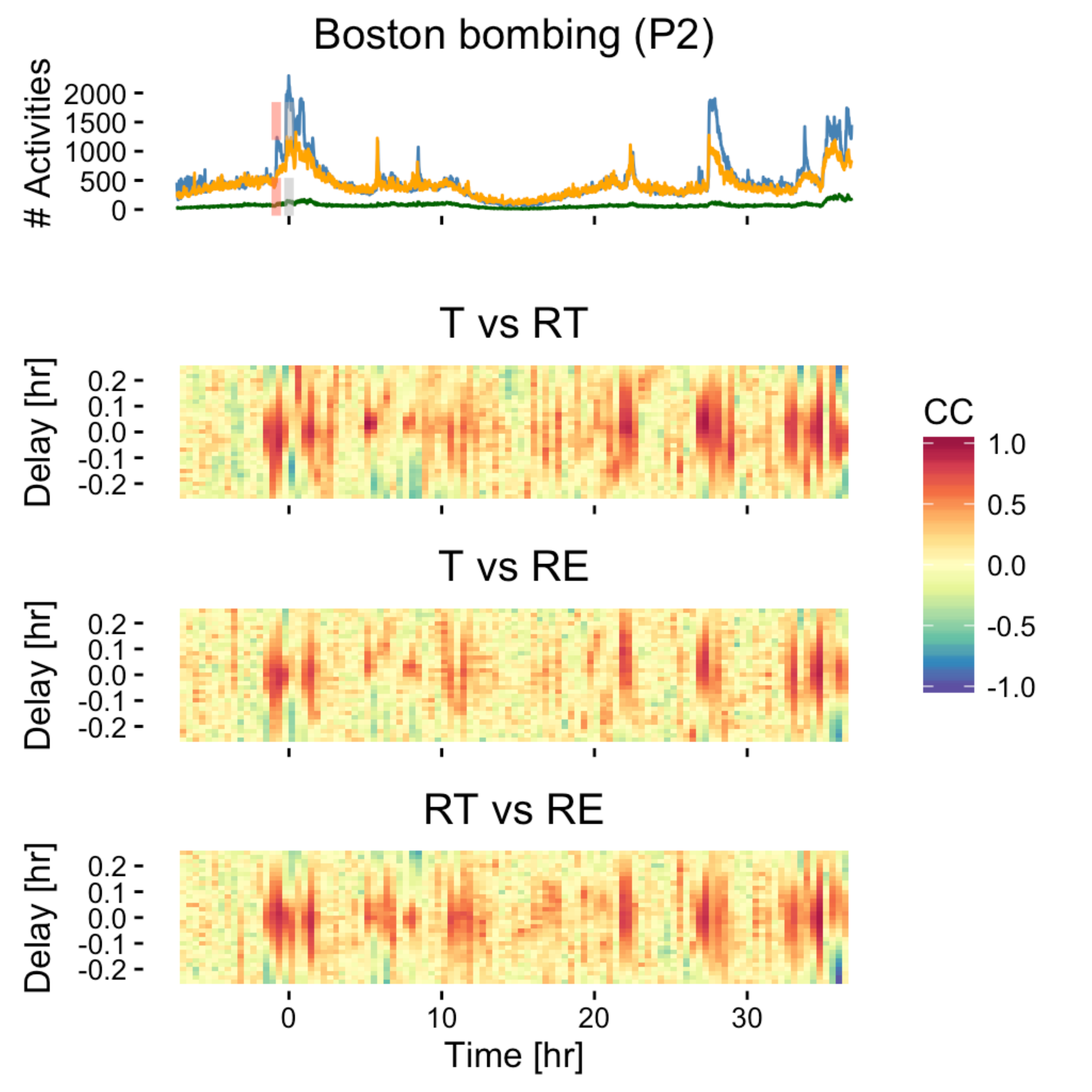}
\caption{{\bf Correlated activity during collective attention.} From top to bottom: i) volume per minute of non-social (orange) and social activities (light blue for retweets, dark green for replies); ii) cross-correlation, color coded, as a function of temporal delay ($y$ axis) and natural time ($x$ axis) between different activities.}
\label{fig:demultiplex}
\end{figure*}

\begin{figure*}
\centering
\includegraphics[width=16cm]{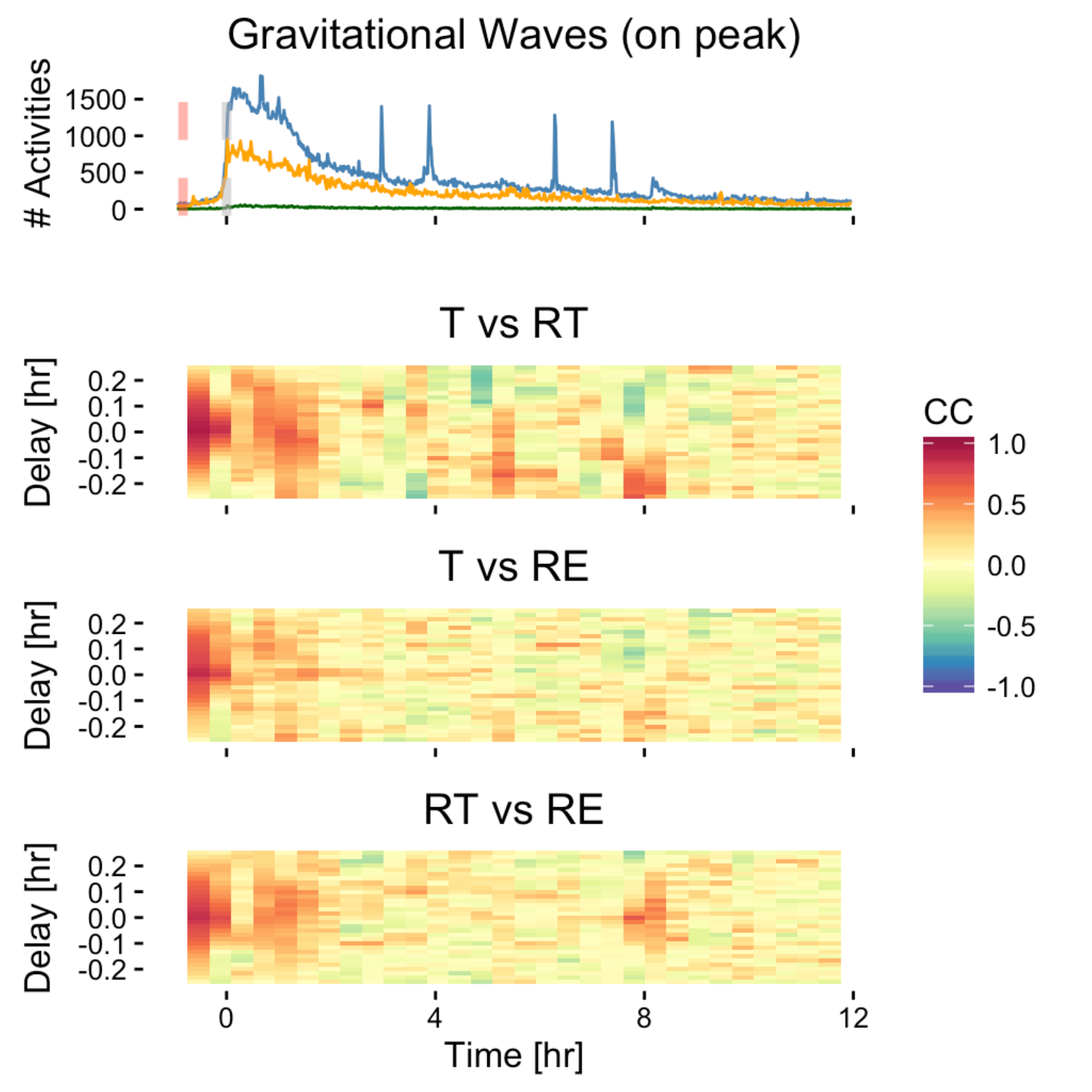}
\caption{{\bf Correlated activity during collective attention.} From top to bottom: i) volume per minute of non-social (orange) and social activities (light blue for retweets, dark green for replies); ii) cross-correlation, color coded, as a function of temporal delay ($y$ axis) and natural time ($x$ axis) between different activities.}
\label{fig:demultiplex}
\end{figure*}

\begin{figure*}
\centering
\includegraphics[width=16cm]{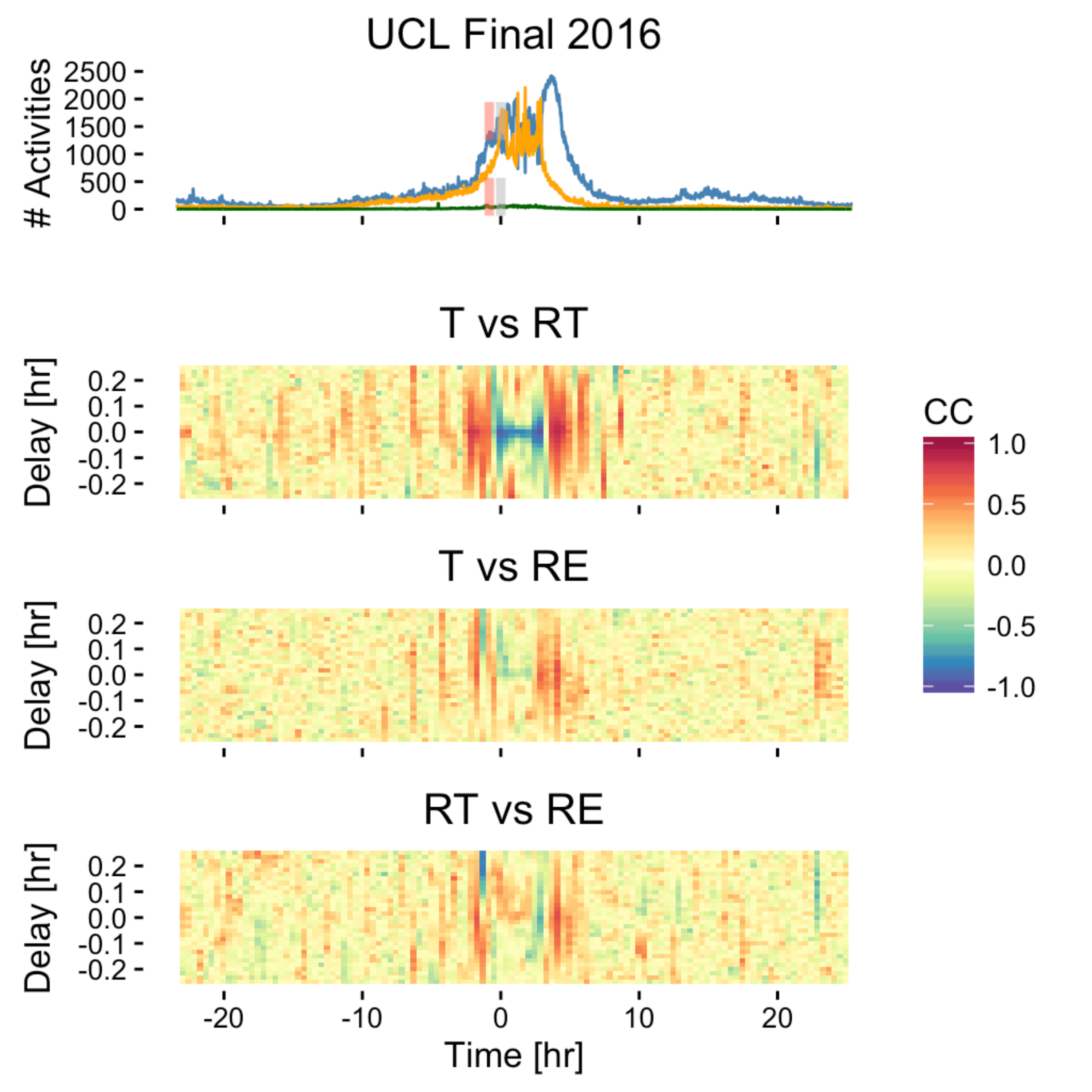}
\caption{{\bf Correlated activity during collective attention.} From top to bottom: i) volume per minute of non-social (orange) and social activities (light blue for retweets, dark green for replies); ii) cross-correlation, color coded, as a function of temporal delay ($y$ axis) and natural time ($x$ axis) between different activities.}
\label{fig:demultiplex}
\end{figure*}

\begin{figure*}
\centering
\includegraphics[width=16cm]{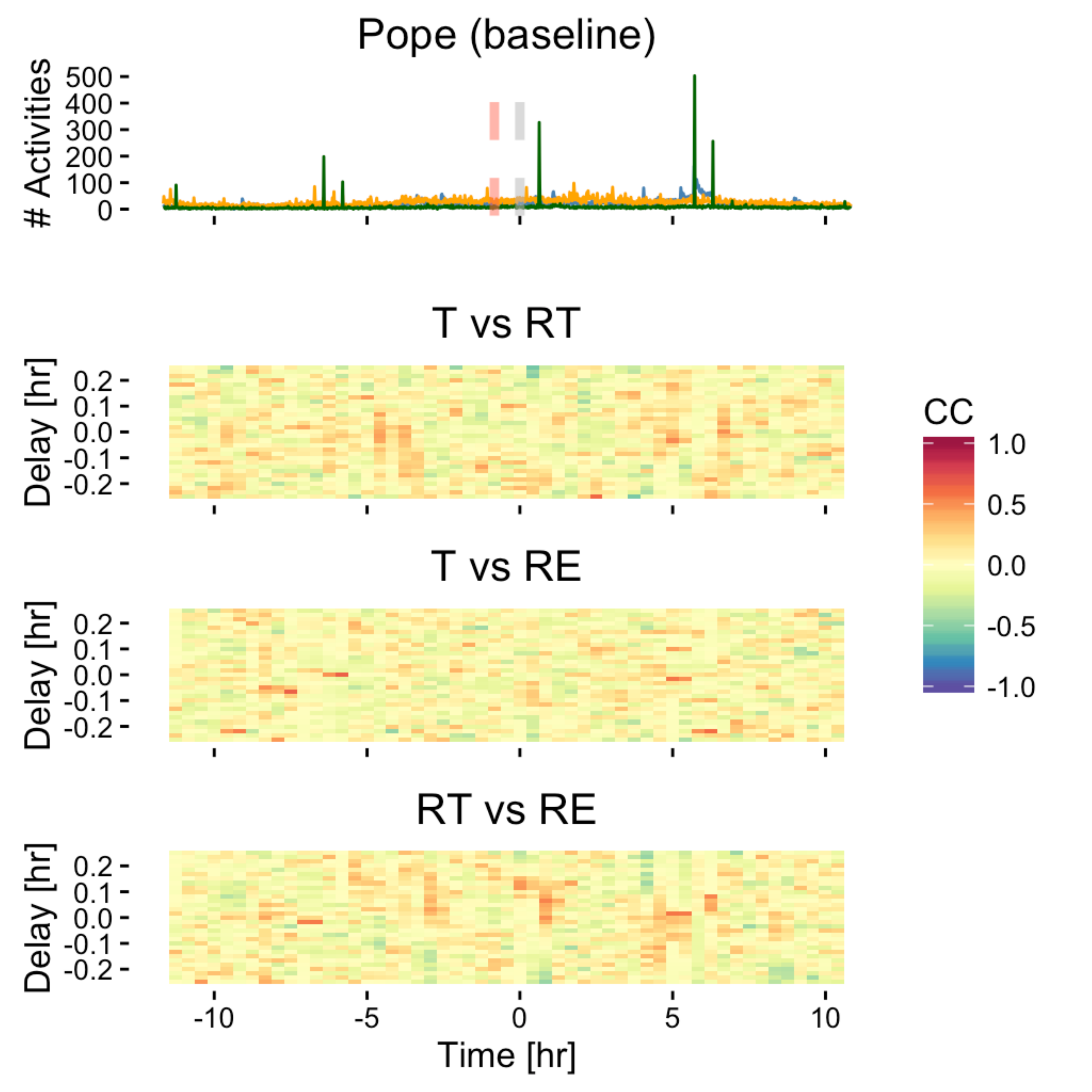}
\caption{{\bf Correlated activity during collective attention.} From top to bottom: i) volume per minute of non-social (orange) and social activities (light blue for retweets, dark green for replies); ii) cross-correlation, color coded, as a function of temporal delay ($y$ axis) and natural time ($x$ axis) between different activities.}
\label{fig:demultiplex}
\end{figure*}

\begin{figure*}
\centering
\includegraphics[width=16cm]{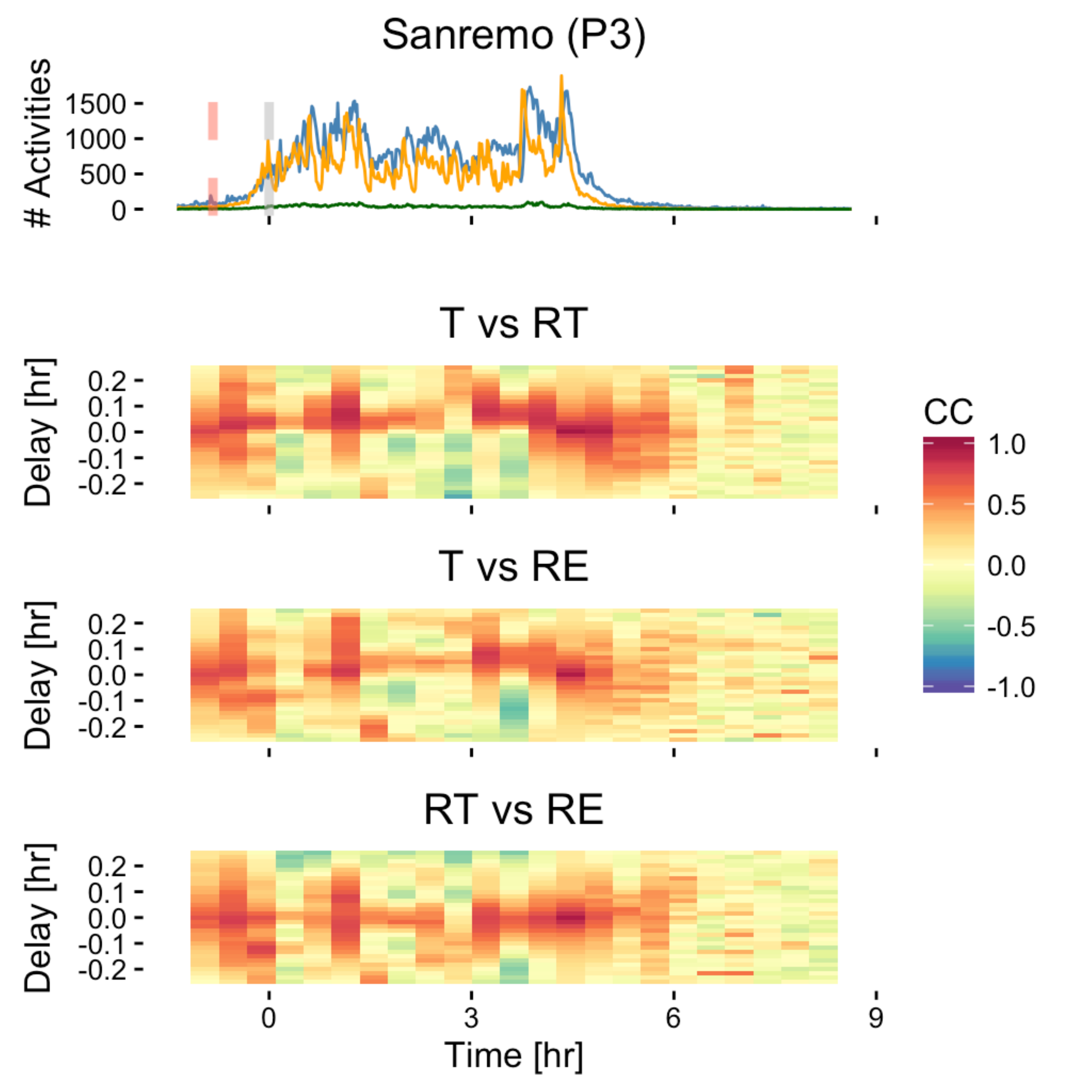}
\caption{{\bf Correlated activity during collective attention.} From top to bottom: i) volume per minute of non-social (orange) and social activities (light blue for retweets, dark green for replies); ii) cross-correlation, color coded, as a function of temporal delay ($y$ axis) and natural time ($x$ axis) between different activities.}
\label{fig:demultiplex}
\end{figure*}

\end{document}